\documentclass[letter]{aa}

\usepackage{graphicx}
\usepackage{txfonts}

\usepackage[colorlinks=true, citecolor=blue, linkcolor=blue, urlcolor=blue]{hyperref}

\definecolor{revcolor}{rgb}{0.70,0.00,0.00}
\newcommand{\rev}[1]{#1}

\begin{document}

   \title{Constraining the gamma-ray efficiency of LINER outflows with Fermi-LAT and MEGARA}

    \titlerunning{Gamma-ray efficiency of LINER outflows}

\author{Alberto Dom\'inguez\inst{1,2}\fnmsep\thanks{alberto.d@ucm.es}
        \and
        Alejandra León\inst{1}\fnmsep\thanks{marleo07@ucm.es}
        \and 
        Adithiya Dinesh\inst{1,2}
        \and
        Armando Gil de Paz\inst{2,3}
        }

\institute{Department of EMFTEL, Universidad Complutense de Madrid, E-28040 Madrid, Spain 
           \and
           Instituto de Física de Partículas y del Cosmos (IPARCOS), Universidad Complutense de Madrid, E-28040 Madrid, Spain
           \and
           Departamento de F\'{\i}sica de la Tierra y Astrof\'{\i}sica, Facultad de Ciencias F\'{\i}sicas, Plaza de Ciencias 1, Universidad Complutense de Madrid, E-28040 Madrid, Spain}
           
\date{Received March 10, 2026}

\abstract
{Low-Ionization Nuclear Emission-line Regions (LINERs) commonly host ionized gas outflows. Their role as high-energy particle accelerators is currently debated, especially following the recent very-high-energy $\gamma$-ray detection of NGC~4278, which showed extreme radiative efficiencies that challenged standard shock-driven emission models.}
{We aim to empirically determine the $\gamma$-ray radiative efficiency of a local sample of LINERs to test whether their high-energy emission can be powered by their extended ionized outflows, or whether compact nuclear jets are required.}
{We combined spatially resolved optical integral-field kinematics from the {\it Multi-Espectrógrafo en GTC de Alta Resolución para Astronomía} at the Gran Telescopio de Canarias, yielding the kinetic powers of ionized outflows ($\dot{E}_{\text{OF}}$), with 17 years of \textit{Fermi}-Large Area Telescope observations to derive 0.05--500~GeV luminosities or 95\% confidence upper limits. We constructed an optical/$\gamma$ diagram placing our sample in the context of archetypal starbursts (M82, NGC~253) and radio galaxies (Centaurus~A, M87).}
{We present the first empirical upper limits on the $\gamma$-ray radiative efficiency of LINER outflows as a population. Our likelihood analysis yields no formal detections ($TS \ge 16$) for the LINERs in our sample. The most physically constraining limit is found for the radio-loud LINER NGC~1052, where the maximum efficiency is restricted to $\eta < 41\%$. For the rest of the sample, the \textit{Fermi}-LAT upper limits generally lie well above the kinetic power of the ionized outflows ($\eta \gg 100\%$). While three sources show marginal hints of emission ($9 < TS < 16$), they remain below the discovery threshold.}
{We conclude that the extended ionized outflows in LINERs are highly inefficient high-energy particle accelerators, analogous to starburst superwinds. These sample-level constraints demonstrate that extreme putative efficiencies cannot be sustained by the ionized outflows alone, favoring a compact nuclear jet origin for the most efficient $\gamma$-ray emitting LINERs.}


   \maketitle
\nolinenumbers

\section{Introduction}
\label{sec:intro}

Low-Ionization Nuclear Emission-line Regions \citep[LINERs;][]{Heckman1980} represent the dominant active galactic nuclei (AGN) population in the local Universe \citep[e.g.,][]{Ho2008,Masegosa2011}. While characterized by radiatively inefficient accretion flows \citep[e.g.,][]{Yuan2014}, high-resolution optical and radio observations reveal that many host complex feedback mechanisms, including compact radio jets \citep[e.g.,][]{Nagar2005} and extended ionized gas outflows \citep[e.g.,][]{Cazzoli2018, HermosaMunoz2020}. Recent integral-field spectroscopy (IFS) surveys using the Gran Telescopio Canarias (MEGARA/GTC) have resolved these outflows, enabling precise measurements of their mass outflow rates and kinetic power \citep[$\dot{E}_{\text{OF}}$;][]{HermosaMunoz2024}. A critical, unresolved question is whether the shocks associated with these outflows can efficiently accelerate cosmic rays to relativistic energies, thereby powering high-energy $\gamma$-ray emission via hadronic or leptonic processes.

Recent \textit{Fermi}-LAT studies have placed upper limits on the $\gamma$-ray emission from the general LINER and low-luminosity AGN populations \citep[LLAGN,][]{deMenezes2020, Karwin2026}. While these works characterize the high-energy properties of the population globally, highlighting the significant contribution of star-formation activity, the specific $\gamma$-ray efficiency of spatially resolved, ionized outflows remains empirically unconstrained.

Extended shock fronts in starburst galaxies (e.g., M82, NGC~253) are known to act as \textit{in situ} particle accelerators, producing detectable GeV--TeV $\gamma$-ray emission \citep[e.g.,][]{Acero2009, Abdo2010a_M82}. Conversely, the recent very-high-energy \citep[VHE;][]{Cao2024} $\gamma$-ray detection of the LINER NGC~4278 poses a serious energetic tension: the observed $\gamma$-ray luminosity exceeds the kinetic power of its ionized outflow by several orders of magnitude. This requires an implausibly high radiative efficiency ($\eta \gg 100\%$), favoring a compact, jet-driven origin \citep[e.g.,][]{Bronzini2024, Dominguez2025}. 

To determine whether the extreme radiative efficiency of NGC~4278 is a unique anomaly or a common trait, we present a systematic search for GeV emission in a well-defined sample of local LINERs with precise optical outflow energetics \citep{HermosaMunoz2024}. Combining 17 years of \textit{Fermi}-LAT observations with these kinematic data, we construct a diagnostic diagram to constrain LINER outflow acceleration efficiencies against archetypal starbursts and radio galaxies.

\section{Sample selection and multiwavelength energetics}
\label{sec:sample}
Our primary sample comprises six local LINERs selected from a high-resolution optical IFS survey performed with MEGARA at the GTC \citep{HermosaMunoz2022, HermosaMunoz2024}, chosen because they possess spatially resolved ionized gas outflows measured with a uniform methodology, which is necessary to obtain precise estimates of their kinetic power ($\dot{E}_{\text{OF}}$). Five of them (NGC~3226, NGC~3245, NGC~4278, NGC~4438, and NGC~4750) represent the radio-quiet to moderately radio-emitting LINER population, typically featuring bubble-like or complex nuclear outflows. The sixth target, NGC~1052, is a well-known radio-loud LINER that hosts a prominent sub-parsec twin jet \citep{Kadler2004}, providing an internal reference for jet-driven feedback; its ionized outflow is aligned with, and most likely driven by, this jet \citep{Cazzoli2022}. Jet-driven feedback has also been reported for NGC~4438 \citep{Puig2026}. While other types of outflows exist in LINERs \citep[e.g., ultra-fast outflows or molecular outflows,][]{Ajello2021}, they currently lack the uniform, high-resolution spatial constraints required for this specific optical-to-$\gamma$-ray energetic comparison.

A central parameter in our study is the kinetic power of the ionized outflows, $\dot{E}_{\text{OF}}$. We adopt the values derived by \citet{HermosaMunoz2024}, and by \citet{Cazzoli2022} for NGC~1052, who isolated the outflow kinematics through multi-component Gaussian fitting of the optical emission lines, identifying the outflow as a broad, blueshifted component. The ionized outflow mass was derived from the H$\alpha$ luminosity (H$\beta$ for NGC~4278) and the electron density estimated from the [S~II]\,$\lambda\lambda6716,6731$\AA\ doublet ratio. Mass outflow rates were then computed assuming either a bi-conical wind or an expanding shell, depending on the resolved morphology, and combined with the maximum velocity and velocity dispersion ($\dot{E}\propto\dot{M}(V^2+3\sigma^2)$). We stress that $\dot{E}_{\text{OF}}$ measures only the kinetic energy flux of the ionized phase \rev{of the outflow}: it neither identifies the driving mechanism nor includes radiative heating, thermal energy, or neutral and molecular outflows. It is therefore a lower limit to the total mechanical energy available for \textit{in situ} particle acceleration, making the efficiencies derived below conservative upper limits. For NGC~4278, \citet{HermosaMunoz2024} report $\dot{E}_{\text{OF}}$ as a lower limit because the MEGARA field of view may not encompass the full outflow; we treat it accordingly. Uncertainties in $\dot{E}_{\text{OF}}$ ($\sim50\%$ for NGC~4438) are typical of IFS studies and are dominated by the assumed electron density, projection effects, and filling factor.

To contextualize the $\gamma$-ray efficiency of our LINERs, we include four well-studied local galaxies: the starbursts M82 and NGC~253, whose emission is mainly attributed to hadronic interactions ($pp$ collisions) of cosmic rays accelerated in shock-driven superwinds \citep[e.g.,][]{Acero2009,Abdo2010a_M82}, and the radio galaxies Centaurus~A and M87, whose relativistic jets dominate their broadband emission \citep[e.g.,][]{Abdo2009_M87,Abdo2010b_CenA}. Their kinetic powers are not derived from H$\alpha$ or measured on the same scales as our MEGARA data: for the starbursts they correspond to the kpc-scale superwind, modeled from diffuse X-ray and optical emission, and for the radio galaxies to the time-averaged jet power \rev{($\dot{E}_{\text{jet}}$)} inferred from X-ray cavities and shells on $\sim10$--100~kpc scales (references in Table~\ref{tab:results}). The resulting systematic uncertainty along the horizontal axis of Fig.~\ref{fig:diagnostic}, up to a factor of a few, is small compared with the \rev{more than five orders of magnitude in kinetic power covered by Fig.~\ref{fig:diagnostic}}. We also homogenized the literature $\gamma$-ray luminosities to our \rev{0.05--500~GeV} band by extrapolating values reported over narrower ranges (e.g., 0.1--5~GeV for M82 and NGC~253) using the published photon index for each source, increasing them by $\sim61\%$ on average.

\section{\textit{Fermi}-LAT data analysis}
\label{sec:analysis}

We searched for high-energy emission from the MEGARA LINER sample by analyzing 17 years of \textit{Fermi}-LAT Pass 8 data \citep[August 2008 to November 2025,][]{Atwood2009,Atwood2013_Pass8,Bruel2018_Pass8Selection} using the \texttt{Fermipy} package \citep[v1.4.0;][]{Wood2017} and \texttt{Fermitools} (v2.5.1)\footnote{\url{https://fermi.gsfc.nasa.gov/ssc/data/analysis/documentation/}}. For each galaxy, we defined a $10^\circ$ Region of Interest (ROI) and adopted a joint likelihood analysis with energy-dependent PSF event types (PSF3 for 50--100~MeV, PSF2--3 for 100--300~MeV, PSF1--3 for 300~MeV--1~GeV, and PSF0--3 for 1--500~GeV) to optimize angular resolution while retaining low-energy sensitivity. At the low redshifts of our sample, $\gamma$-ray attenuation by the extragalactic background light is negligible \citep[e.g.,][]{SaldanaLopez2021}. Our baseline background model incorporated all 14-year 4FGL-DR4 sources within $15^\circ \times 15^\circ$ \citep{Abdollahi2022, Ballet2023}, the standard Galactic diffuse template (\texttt{gll\_iem\_v07.fits}) with free normalization and spectral index, and the corresponding PSF-dependent isotropic templates (e.g., \texttt{iso\_P8R3\_SOURCE\_V3\_PSF3\_v1.txt})\footnote{\url{https://fermi.gsfc.nasa.gov/ssc/data/access/lat/BackgroundModels.html}}.

Following recent LLAGN studies \citep{Karwin2026}, our binned maximum likelihood analysis applied a background modeling threshold of $TS \ge 16$ and a formal target discovery threshold of $TS \ge 25$ ($\approx 5\sigma$), where the test statistic $TS = 2(\ln\mathcal{L}_1 - \ln\mathcal{L}_0)$ compares the likelihood of the model with and without the target \citep{Mattox1996}. Our analysis yielded no significant target excesses, even at the lower $TS \ge 16$ level. Consequently, we computed 95\% confidence level upper limits on the photon and energy fluxes via a profile likelihood method. For the upper-limit calculations, the target photon index was frozen at $\Gamma = 2.2$, which is a typical average value adopted for faint extragalactic sources \citep[e.g.,][]{Karwin2026}. Varying $\Gamma$ between 1.8 and 2.5 altered the limits by at most $\sim 20\%$, confirming this assumption does not drive our conclusions. The energy-flux upper limits were then converted into 0.05--500~GeV luminosities. The event selection, the treatment of the free parameters, and the derivation of the upper limits are detailed in Appendix~\ref{app:lat}.

\section{Results and discussion}
\label{sec:results}

\subsection{Gamma-ray luminosities and the diagnostic diagram}

Our likelihood analysis of the 17-year \textit{Fermi}-LAT data yielded no significant detections ($TS \ge 16$) at the positions of the six LINERs in our sample, although three of them show marginal hints of emission ($9 < TS < 16$; Sect.~\ref{sec:marginal}). Consequently, we calculated 95\% confidence level upper limits on the 0.05--500~GeV luminosities for the entire sample. Table~\ref{tab:results} summarizes these measurements.

\begin{table}
\caption{Multiwavelength energetics and \textit{Fermi}-LAT observations
(0.05--500 GeV) for the MEGARA LINER sample and literature comparison sources.}
\label{tab:results}
\centering
\scriptsize
\setlength{\tabcolsep}{3pt}
\begin{tabular}{lcccccc}
\hline\hline
Source & $D$ & \rev{$\dot{E}_{\text{OF}}$/$\dot{E}_{\text{jet}}$} & $TS$ & $F_\gamma$ & $L_\gamma$ & $\eta$ \\
 & (Mpc) & (erg s$^{-1}$) & & (ph cm$^{-2}$s$^{-1}$) & (erg s$^{-1}$) & (\%) \\
\hline
NGC 1052 & 19.4 & $(8.8\pm3.5)\times10^{40}$ & 1.6  & $1.99\times10^{-9}$  & $3.63\times10^{40}$ & $<41$ \\
NGC 3226 & 23.6 & $(1.7\pm0.6)\times10^{40}$ & 14.1 & $3.78\times10^{-9}$  & $1.01\times10^{41}$ & $<590$ \\
NGC 3245 & 20.9 & $(1.2\pm0.7)\times10^{40}$ & 0.3  & $1.78\times10^{-9}$  & $3.75\times10^{40}$ & $<310$ \\
NGC 4278 & 16.4 & $>7.2\times10^{37}$\tablefootmark{a} & 15.1 & $1.19\times10^{-9}$  & $1.54\times10^{40}$ & $<2.1\times10^{4}$ \\
NGC 4438 & 17.0 & $(2.7\pm1.4)\times10^{38}$ & 0.0  & $3.09\times10^{-10}$ & $4.31\times10^{39}$ & $<1600$ \\
NGC 4750 & 26.1 & $(7.5\pm2.7)\times10^{39}$ & 12.2 & $2.37\times10^{-9}$  & $7.81\times10^{40}$ & $<1040$ \\
\hline
\multicolumn{7}{c}{Literature Comparison Sample} \\
\hline
M82     & 3.6  & $3.1\times10^{42}$ & -- & -- & $2.3\times10^{40}$ & 0.74 \\
NGC 253 & 3.9  & $1.5\times10^{42}$ & -- & -- & $1.3\times10^{40}$ & 0.87 \\
M87     & 16.0 & $2.4\times10^{43}$ & -- & -- & $6.3\times10^{41}$ & 2.6 \\
Cen A   & 3.7  & $1.0\times10^{43}$ & -- & -- & $4.2\times10^{39}$ & 0.042 \\
\hline
\end{tabular}
\tablefoot{Distances ($D$) are from the literature. $\dot{E}_{\text{OF}}$ values
are from \citet{HermosaMunoz2024}, except NGC~1052
(\citealt{Cazzoli2022}). For the comparison sample, the
tabulated power is the superwind power $\dot{E}_{\text{OF}}$ (M82, NGC~253) or
\rev{the time-averaged jet power $\dot{E}_{\text{jet}}$} (M87, Cen~A;
Sect.~\ref{sec:sample}). No source reached formal
$\gamma$-ray detection ($TS\ge25$); $\Gamma=2.2$ was fixed and 95\%
confidence upper limits are reported for $F_\gamma$ and $L_\gamma$. Three
sources show marginal emission ($9<TS<16$). $\eta =
(L_\gamma/\rev{\dot{E}})\times100\%$; values for our targets are upper
limits because $L_\gamma$ is an upper limit and
$\dot{E}_{\text{OF}}$ a lower limit to the total mechanical budget
(Sect.~\ref{sec:sample}). Literature values were homogenized to the
0.05--500~GeV range.
\tablefoottext{a}{Lower limit: outflow may extend beyond the MEGARA field;
energetics are based on H$\beta$ instead of H$\alpha$
\citep{HermosaMunoz2024}.}}
\tablebib{Literature comparison sample:
M82 (\citealt{Strickland2009,deCeaDelPozo2009,Abdo2010a_M82});
NGC~253 (\citealt{Romero2018,Acero2009});
M87 (\citealt{Forman2005,Abdo2009_M87});
Cen~A (\citealt{Croston2009,Abdo2010b_CenA,Brown2017}).}
\end{table}

\begin{figure}[ht]
\centering
 \includegraphics[width=\columnwidth]{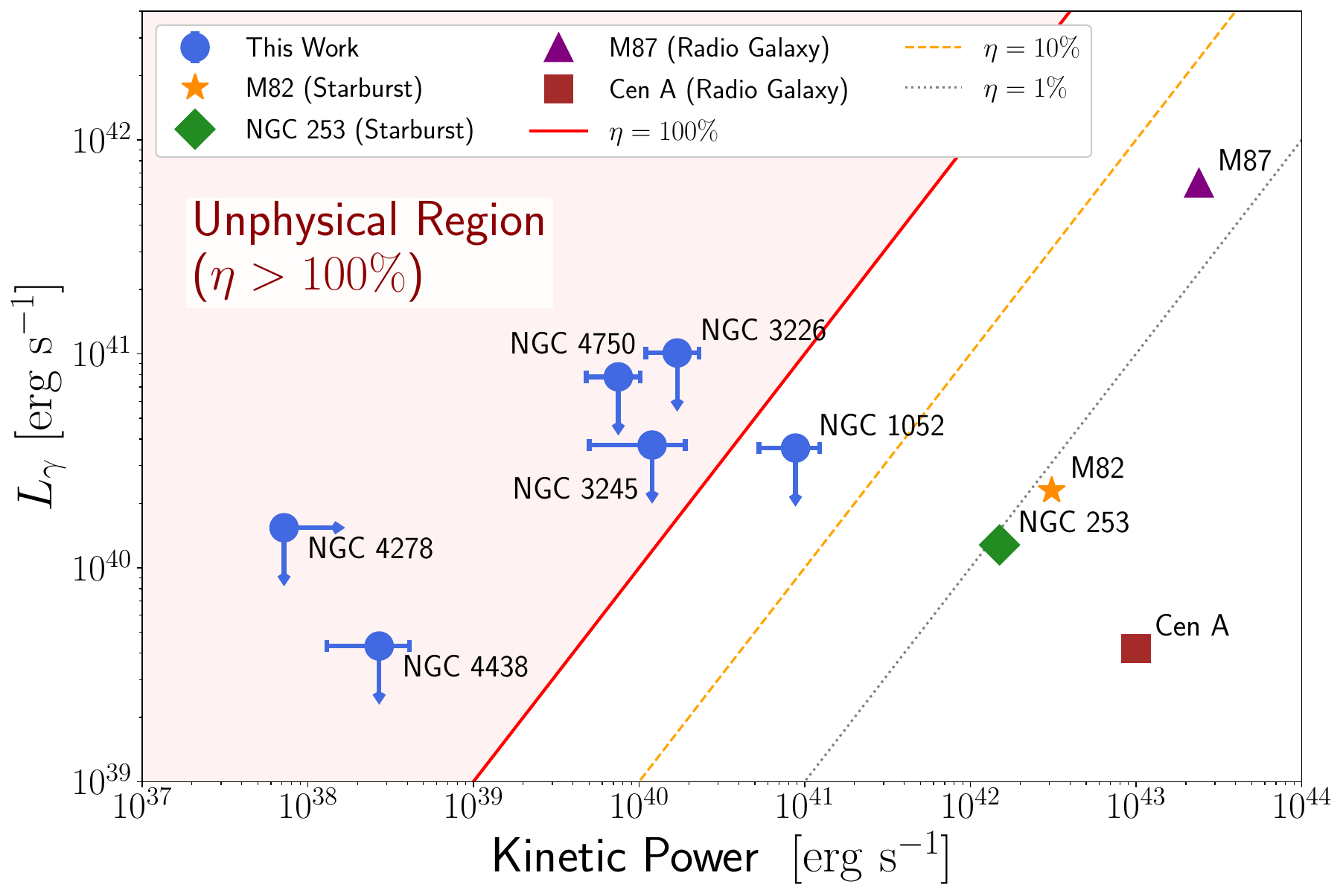}
\caption{Diagnostic diagram comparing $L_\gamma$ and kinetic power. For the MEGARA LINER sample and the starburst galaxies (M82, NGC~253), the x-axis represents the kinetic power of the extended ionized outflows and of the star-formation-driven superwind, respectively. For the radio galaxies (Centaurus~A, M87), the x-axis represents the time-averaged kinetic power of their jets. Literature $\gamma$-ray luminosities have been homogenized to the 0.05--500~GeV band for consistency. Diagonal lines represent contours of constant putative efficiency ($\eta$). Downward-pointing arrows indicate 95\% confidence level upper limits on $L_\gamma$; the rightward arrow on NGC~4278 indicates that its outflow kinetic power is a lower limit \citep{HermosaMunoz2024}.}
\label{fig:diagnostic}
\end{figure}

To evaluate the physical origin of the high-energy emission, we compare $L_\gamma$ and $\dot{E}_{\text{OF}}$ in the diagnostic diagram of Fig.~\ref{fig:diagnostic}, which includes the MEGARA sample and literature comparison objects. We define the efficiency as $\eta=(L_\gamma/\dot{E}_{\text{OF}})\times100\%$.

The datasets probe different scales ($\lesssim 1.5$~kpc for MEGARA, tens of kpc for the effective LAT aperture). As discussed in Appendix~\ref{app:scales}, both mismatches are conservative, so the $\eta$ values in Table~\ref{tab:results} are robust upper limits on the true radiative efficiency of the ionized outflows.

\subsection{The radio-loud control: NGC 1052}
The most physically constraining limit in our sample is found for the radio-loud LINER NGC~1052, where the maximum efficiency is restricted to $\eta < 41\%$. NGC~1052 is known to host a prominent sub-parsec twin radio jet with a kinetic power exceeding $10^{43}$ erg s$^{-1}$ \citep{Kadler2004}. The fact that its $\gamma$-ray emission is constrained below $41\%$ of its much weaker ionized outflow power \citep[$\dot{E}_{\text{OF}} \approx 8.8 \times 10^{40}$ erg s$^{-1}$;][]{Cazzoli2022} shows that even in a known jet-dominated system the extended ionized outflow is a highly inefficient particle accelerator, making NGC~1052 the ideal internal control for distinguishing jet-driven from outflow-driven high-energy emission.

\subsection{Marginal hints of emission}
\label{sec:marginal}
Three sources in our sample (NGC~3226, NGC~4278, and NGC~4750) showed marginal hints of $\gamma$-ray emission ($9 < TS < 16$, corresponding to $\sim 3\text{--}4\sigma$). Because these targets lie in somewhat crowded fields, we carefully assessed potential spatial contamination: NGC~4278 lies within $\sim$\,1$^\circ$ of several bright Coma cluster blazars, and the unassociated source 4FGL~J1242.9+7315 ($TS \sim 170$) only $0.65^\circ$ from NGC~4750. All are included in our background model, and our energy-dependent PSF selections reach an angular resolution of $\sim$\,0.1$^\circ$--$0.2^\circ$ above 1~GeV, comfortably resolving these separations. The TS maps (available upon request, see Data availability) confirm that the marginal excesses ($TS = 15.1$ and $12.2$, respectively) are spatially coincident with the optical positions of the galaxies and distinct from the neighboring sources. For NGC~3226, however, the LAT cannot separate the LINER from its interacting Seyfert~1 companion NGC~3227, only $\sim 2\arcmin$ away, so its excess and upper limit apply to the pair as a whole.

Having ruled out spatial contamination, we consider the physical nature of these excesses. Although we formally report NGC~4278 ($TS = 15.1$, $\sim 3.9\sigma$) as an upper limit, to keep a consistent threshold across our predefined sample, the independent VHE detections \citep{Bronzini2024, Cao2024} make it plausible that this excess is real baseline emission, since integrating over 17 years dilutes transient jet emission into a low-significance steady signal. Furthermore, both NGC~3226 and NGC~4750 possess compact radio cores. While the 17-year integrated \textit{Fermi}-LAT signal-to-noise ratio is too low to extract meaningful light curves via a blind short-timescale search, the presence of these $\sim 3\sigma$ excesses is highly suggestive of unresolved, variable jet activity rather than steady, outflow-driven emission.

\subsection{The efficiency limits of extended LINER outflows}

The location of a source on the $L_\gamma$--$\dot{E}_{\text{OF}}$ plane measures its particle acceleration efficiency. The starbursts M82 and NGC~253 populate the low-efficiency region ($\eta \lesssim 1\%$): comparing the supernova-driven mechanical power of M82 \citep[$\dot{E}_{\text{OF}} \sim 3.1 \times 10^{42}$ erg s$^{-1}$;][]{Strickland2009, deCeaDelPozo2009} with its homogenized luminosity ($L_{\gamma} \sim 2.3 \times 10^{40}$ erg s$^{-1}$) gives $\eta \approx 0.7\%$.

For most of our radio-quiet LINERs the upper limits sit well above the measured ionized outflow powers ($\eta \gg 100\%$, Table~\ref{tab:results}); at starburst-like efficiencies ($\eta \sim 0.7\%$) their expected luminosities would fall three to four orders of magnitude below current GeV thresholds. Our non-detections are therefore consistent with extended LINER outflows acting as standard, low-efficiency accelerators.

\subsection{The requirement for compact emission}
This is crucial for interpreting the VHE $\gamma$-ray detection of NGC~4278 \citep{Cao2024}, which shows an efficiency $\eta \lesssim 2 \times 10^{4}\%$ when its $\gamma$-ray luminosity is compared to the kinetic power of its outflow \citep[as pointed out by][]{Dominguez2025}. Because that power is a lower limit \citep{HermosaMunoz2024}, we note that the outflow would have to be more than two orders of magnitude more energetic than measured within the MEGARA field of view for $\eta$ to fall below $100\%$, which is implausible for a compact nuclear outflow. The segregation of NGC~4278, M87, and Cen~A from the outflow-dominated region of the diagram indicates that efficient $\gamma$-ray emission requires a compact nuclear jet. M87 is particularly instructive, since it is itself classified as a LINER \citep{Ho2008}: its position in Fig.~\ref{fig:diagnostic} is set by the large power channeled through its jet, \rev{plotted here as $\dot{E}_{\text{jet}}$ rather than as an outflow power,} while its efficiency ($\eta \approx 2.6\%$) remains comparable to that of the starbursts. It is thus the power available to a compact jet, rather than an anomalous efficiency, that makes a LINER detectable at GeV energies. This is consistent with spectral modeling of the subthreshold LLAGN population, which shows that advection-dominated accretion flows alone underpredict the GeV emission. Explaining the emission requires additional components: star formation for the general population or compact radio jets for the most efficient systems \citep{Karwin2026}. Unlike extended outflows, relativistic jets channel accretion energy into collimated regions, driving rapid particle acceleration. The two mechanisms can be distinguished by their $\gamma$-ray properties: jet-driven emission shows harder spectra and short-term variability, whereas star formation produces steady, softer emission.

\section{Summary and Conclusions}
\label{sec:conclusions}
We have investigated the high-energy radiative efficiency of local LINERs by comparing 17 years of Fermi-LAT observations with the kinetic power of their spatially resolved ionized outflows measured by MEGARA/GTC, to determine whether these outflows can efficiently accelerate particles to relativistic energies, or if a compact, jet-driven origin is required. Our main conclusions are as follows: (1) We report no significant GeV detections ($TS \ge 16$), and derive stringent 95\% confidence level upper limits on the 0.05--500~GeV luminosities of the whole sample; three sources show marginal hints of emission ($9 < TS < 16$). (2) On an $L_\gamma$--$\dot{E}_{\text{OF}}$ diagnostic diagram, the upper limits generally lie well above the kinetic power of the ionized outflows ($\eta \gg 100\%$), except for the radio-loud LINER NGC~1052 ($\eta < 41\%$). Because a detection powered solely by the ionized outflow would require implausibly high efficiencies, our non-detections are physically consistent with LINER outflows operating as standard, low-efficiency particle accelerators ($\eta \lesssim 1\%$), analogous to starburst superwinds. (3) The extreme efficiency inferred for the VHE-detected LINER NGC~4278 ($\eta \lesssim 2 \times 10^{4}\%$) cannot be sustained by its ionized outflow, even allowing for its kinetic power being a lower limit; our sample-level constraints demonstrate that this is not a general property of LINER outflows. We conclude that when a LINER shows highly efficient $\gamma$-ray emission, the particle acceleration is consistent with being driven by a compact nuclear jet.

Ultimately, while ionized outflows are ubiquitous in LINERs, they are generally inefficient high-energy emitters: at starburst-like efficiencies ($\eta \sim 0.7\%$), an outflow would need $\dot{E}_{\text{OF}} \gtrsim 1.4 \times 10^{41}$ erg s$^{-1}$ to reach a detectable $\gamma$-ray luminosity of $\sim 10^{39}$ erg s$^{-1}$. Future observations with the Cherenkov Telescope Array Observatory (CTAO) will be crucial to reach these sensitivities, allowing us to map the boundary between outflow-driven and jet-driven feedback at the faint end of the AGN population.

\begin{acknowledgements}
\rev{AD acknowledges support from MCIN/AEI under grant PID2025-171816NB-I00; A. Dinesh under grant PID2022-138132NB-C42; and AGdP under grant PID2022-138621NB-I00 (MCIN/AEI/10.13039/501100011033/FEDER, EU). We thank the anonymous referee for a constructive report.} The \textit{Fermi} LAT Collaboration acknowledges generous ongoing support from a number of agencies and institutes that have supported both the development and the operation of the LAT as well as scientific data analysis. These include the National Aeronautics and Space Administration and the Department of Energy in the United States, the Commissariat \`a l'Energie Atomique and the Centre National de la Recherche Scientifique / Institut National de Physique Nucl\'eaire et de Physique des Particules in France, the Agenzia Spaziale Italiana and the Istituto Nazionale di Fisica Nucleare in Italy, the Ministry of Education, Culture, Sports, Science and Technology (MEXT), High Energy Accelerator Research Organization (KEK) and Japan Aerospace Exploration Agency (JAXA) in Japan, and the K. A. Wallenberg Foundation, the Swedish Research Council and the Swedish National Space Board in Sweden. Additional support for science analysis during the operations phase is gratefully acknowledged from the Istituto Nazionale di Astrofisica in Italy and the Centre National d'\'Etudes Spatiales in France. This work performed in part under DOE Contract DE-AC02-76SF00515.

\end{acknowledgements}

\section*{Data availability}

The \textit{Fermi}-LAT data are publicly available from the FSSC (\url{https://fermi.gsfc.nasa.gov/ssc/}). The optical kinematic parameters and outflow energetics ($\dot{E}_{\text{OF}}$) for the MEGARA LINER sample are available in \citet{HermosaMunoz2024} and \citet{Cazzoli2022}. Intermediate likelihood-analysis products are available from the corresponding author upon reasonable request.

\appendix
\nolinenumbers

\section{Details of the \textit{Fermi}-LAT analysis}
\label{app:lat}

This appendix expands on the methodology of Sect.~\ref{sec:analysis}, so that the analysis can be reproduced by readers outside the $\gamma$-ray community.

\subsection{Event selection and binning}

We used Pass 8 \texttt{SOURCE}-class events (\texttt{evclass}~$=128$, \texttt{P8R3\_SOURCE\_V3} instrument response functions) within $10^\circ$ of each target in the 0.05--500~GeV band, filtered with the standard good-time-interval selection \texttt{(DATA\_QUAL>0)\&\&(LAT\_CONFIG==1)}. The binned analysis used a $10^\circ \times 10^\circ$ ROI with a spatial binning of 0\fdg08 pixel$^{-1}$ and eight logarithmic energy bins per decade. To optimize the angular resolution while retaining low-energy acceptance, the data were split into ten independent components by PSF event type and energy, each with its own isotropic template, and fitted jointly as the product of their likelihoods: PSF3 for 50--100~MeV, PSF2--3 for 100--300~MeV, PSF1--3 for 300~MeV--1~GeV, and PSF0--3 for 1--500~GeV, with maximum zenith angles of $80^\circ$, $90^\circ$, $100^\circ$, and $105^\circ$, respectively, to suppress the contamination from the Earth limb. Energy dispersion was applied to all model components except the isotropic templates, for which it is already included in their empirical derivation.

\subsection{Likelihood fit and free parameters}

The test statistic is defined as $TS = 2(\ln \mathcal{L}_1 - \ln \mathcal{L}_0)$, where $\mathcal{L}_1$ and $\mathcal{L}_0$ are the maximum likelihood values of the model including and excluding the source of interest \citep{Mattox1996}; for a source at a known position with a single free parameter, $\sqrt{TS}$ approximates its significance in Gaussian standard deviations.

The following parameters were left free in the fit: the normalization and spectral index of the Galactic diffuse template; the normalization of each isotropic template; the normalizations of the 4FGL-DR4 sources with $TS \ge 16$ lying within $9^\circ$ of the target, and of those with $TS \ge 500$ within $12^\circ$; and the normalization of the target itself. All remaining parameters, including the spectral shapes of the catalog sources, were fixed at their catalog values. After a first fit, we searched the residuals for sources not included in the catalog and added any excess with $TS \ge 16$ (with a minimum separation of $0\fdg5$) to the model as an additional point source with a power-law spectrum of index $\Gamma = 2.2$, before refitting the ROI.

\subsection{Upper limits}

The targets were modeled as point sources at their optical positions with a power-law spectrum whose photon index was frozen at $\Gamma = 2.2$ and whose normalization was left free (Sect.~\ref{sec:analysis}). Since none of them reached $TS = 25$, we derived 95\% confidence level upper limits on both the photon and the energy flux with the profile likelihood method, scanning the normalization of the target while refitting the remaining free parameters at each step. Table~\ref{tab:results} lists the photon-flux limits; the energy-flux limits $S_E$ were converted into isotropic-equivalent luminosities as $L_\gamma = 4 \pi d_L^2 S_E$, using the distances given in the same table. No beaming correction was applied.

\subsection{Angular scales probed by the two datasets}
\label{app:scales}

The optical and $\gamma$-ray data combined in Fig.~\ref{fig:diagnostic} are collected over very different apertures. The MEGARA field of view (12\farcs5 $\times$ 11\farcs3) corresponds to $\lesssim 1.5$~kpc at the distances of our targets and therefore isolates the nuclear ionized outflow. The $10^\circ$ ROI, by contrast, is used only to build the background model: the aperture over which the target flux is actually measured is set by the LAT PSF, which reaches $\sim 0.1^\circ$--$0.2^\circ$ above 1~GeV, i.e. tens of kpc at 20~Mpc. The LAT limits therefore constrain the \textit{total} GeV output of the galaxy, including the nucleus, the outflow, host star formation and, in interacting systems such as NGC~3226 and NGC~4438, the companion.

Both mismatches act in the same, conservative direction. Any additional emitting component within the LAT aperture implies that the outflow contributes less than the quoted limit, and hence a lower true efficiency. Likewise, any outflow emission falling outside the MEGARA field of view implies that $\dot{E}_{\text{OF}}$ is underestimated, again lowering the true efficiency. The values of $\eta$ quoted in Table~\ref{tab:results} are therefore upper bounds, \rev{which reinforces our conclusions}.

\end{document}